\documentclass[12pt]{iopart}
\usepackage[dvips]{epsfig}

\newcommand{\figwidth}{0.98\columnwidth}
\newcommand{\vect}[1]{\mathbf{#1}}
\newcommand{\phcc}{(C$_4$H$_{12}$N$_2$)(Cu$_2$Cl$_6$){}}
\newcommand{\phxc}{(C$_4$H$_{12}$N$_2$)(Cu$_2$Cl$_{6(1-y)}$Br$_{6y}$){}}
\begin{document}

\title{Formation of gapless triplets in the
bond-doped spin-gap antiferromagnet \phcc.}

\author{V N Glazkov$^{1,2}$, G Skoblin$^{1,2}$, D H\"{u}vonen$^{3,4}$, T S Yankova$^{3,5}$, A Zheludev$^3$}

\address{$^1$ Kapitza Institute for Physical Problems RAS,
Kosygin str. 2, 119334 Moscow, Russia}

\address{$^2$Moscow Institute of Physics and Technology, 141700
Dolgoprudny, Russia}

\address{$^3$Neutron Scattering and Magnetism, Laboratory for
Solid State Physics, ETH Z\"{u}rich, 8006 Z\"{u}rich, Switzerland}

\address{$^4$National Institute of Chemical Physics and Biophysics,
Akadeemia tee 23, 12618 Tallinn, Estonia}

\address{$^5$Chemical Department, M.V.Lomonosov
Moscow State University, Moscow, Russia}

\ead{glazkov@kapitza.ras.ru}

\begin{abstract}

We report results of an electron spin resonance (ESR) study of a
spin-gap antiferromagnet \phcc (nicknamed PHCC) with chlorine ions
partially substituted by bromine. We found that up to 10\% of
nominal doping the contribution of the random defects to the
absorption spectra remains at about 0.1\% per copper ion, almost
the same as in the pure system. Instead, a particular kind of ESR
absorption corresponding to gapless S=1 triplets is observed at
low temperatures in samples with high nominal bromine content
$x\geq5$\%. Increase of bromine concentration also leads to the
systematic broadening of ESR absorption line indicating reduction
of the quasi-particles lifetime.

\end{abstract}


\date{\today}
\submitto{\JPCM}

\section{Introduction.}

Spin-gap antiferromagnets have been actively studied during the
last decades. Because of particular geometry of the spin-spin
interactions (frequently featuring dimer motives) these systems of
strongly exchange coupled spins do not order
antiferromagnetically, but remain in  disordered spin-singlet
($S=0$) state down to the lowest temperatures. Stability of the
singlet state is ensured by an energy gap $\Delta$ of exchange
origin
 separating the ground state from the triplet
($S=1$) excitations. Among  examples of such systems are: a
spin-Peierls magnet CuGeO$_3$ \cite{cugeo3-regnault}, a dimer
magnet TlCuCl$_3$\cite{tlcucl3}, various Haldane magnets (e.g.,
PbNi$_2$V$_2$O$_7$ \cite{pbnio1,pbnio2} or organo-metallic
compound NENP \cite{nenp}) and various spin-ladder systems
\cite{dave-dimpy}.

Application of an external magnetic field leads to the Zeeman
splitting of the triplet states and to lowering of the energy of
one of its substates. This leads to a quantum critical point at a
certain critical field $H_c \simeq \Delta/(g\mu_B)$. This quantum
critical point has been actively discussed recently since it is
formally quivalent to Bose-Einstein condensation
\cite{tlcucl3,giamarchi,zapf-revmodphys}. If a spin subsystem is
three-dimensional, then a field induced antiferromagnetically
ordered phase is formed above the critical field \cite{giamarchi}.

Controlled introduction of  defects is another way to affect
stability of the ground state. Bond-doping, i.e. introduction of
defects that affect interspin couplings but do not affect magnetic
sublattices, is of particular interest as a way to realize
different random coupling models. Theoretical models considering
such systems make interesting predictions, including formation of
a glass-like state\cite{glassy} or a random singlet state
\cite{random,random2}. This mechanism was already exploited in the
study of random-bond systems IPA-CuCl$_3$ \cite{ipa1,ipa2},
Sul-Cu$_2$Cl$_4$ \cite{sul}, and DTN\cite{dtn}.

Recently found organo-metallic compound \phcc (abbreviated as PHCC
for piperazinium hexachlorodicuprate) is a good test model to
study quantum critical behaviour and the effect of impurities. The
pure compound is well characterised by a variety of techniques
including bulk measurements, elastic and inelastic neutron
scattering \cite{stone,broholm-prb64,stone-nature} and electron
spin resonance \cite{glazkov-phcc}. These measurements have
confirmed the existence of an energy gap $\Delta=1.02$ meV,
triplet nature of magnetic excitations and closing of the energy
gap by a magnetic field above approximately 8T. Inelastic neutron
scattering experiments revealed the presence of at least six
relevant exchange couplings\cite{stone}, the exchange couplings
geometry can be envisioned as a 2D set of moderately coupled
spin-ladders with the dominant interaction on the rung of the
ladder and some frustrating couplings.

This system allows to perform a substitution of chlorine ions
mediating the interspin interactions by bromine with the nominal
Br concentration above 10\%. As it was shown by neutronographic
and calorimetric experiments
\cite{prb-dan-doped,prb-dan-doped-magnons,dan-phcc-doped2} doping
results in an increase of the gap and the critical field, as well
as  in a broadening of the excitation spectrum and in a change of
the crossover critical exponent. The magnon damping in the doped
samples turned out to be strongly dependent on the wavevector with
damping rate increasing away from the zone center
\cite{prb-dan-doped-magnons}.

Here we present the results of an electron spin resonance (ESR)
study of the bromine doped PHCC single crystals. High energy
resolution of ESR technique and its ability to distinguish between
different paramagnetic centers allows us to demonstrate that
particular kind of S=1 paramagnetic centers are formed with bond
doping: low temperature ESR response is dominated by resonance
absorption of  gapless triplets. At the same time, increase of ESR
linewidth indicates reduction of lifetime of thermally excited
triplet excitations.

\section{Experimental details and samples characterisation.}

\begin{figure}
  \centering
  \epsfig{file=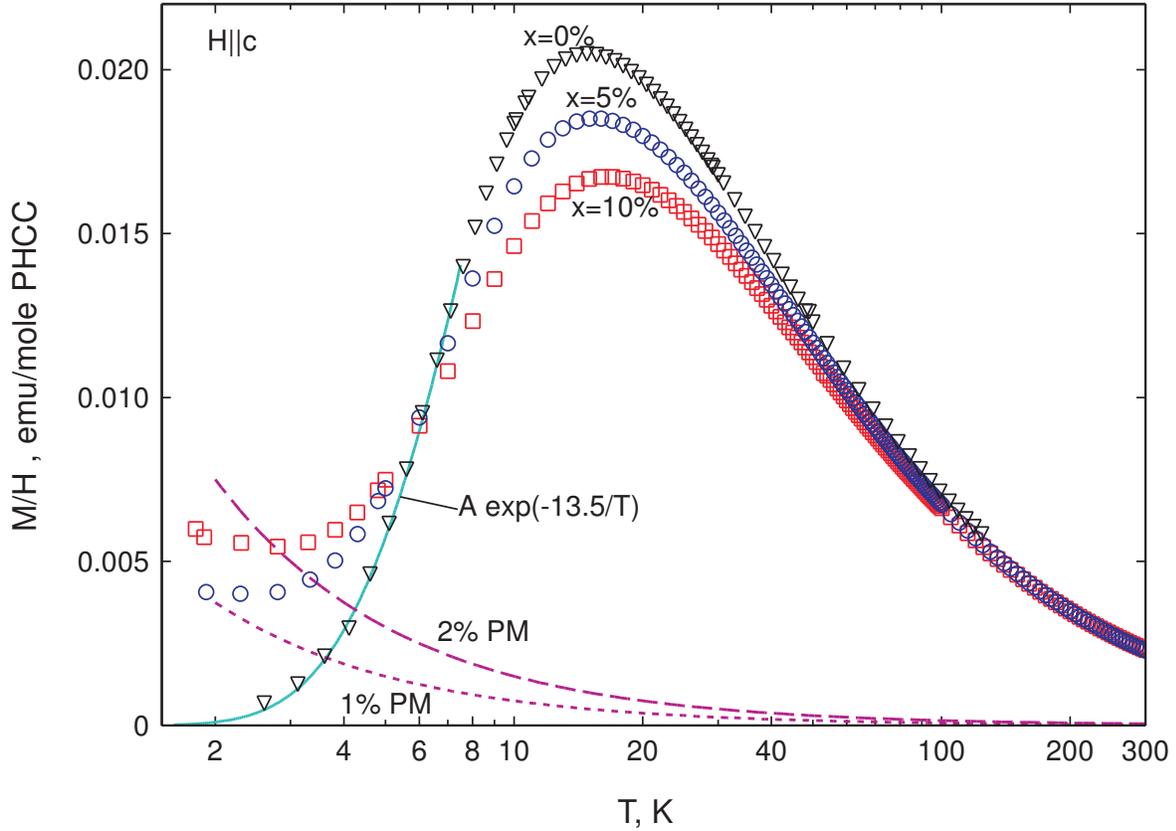, width=\figwidth, clip=}
  \caption{(color online) Temperature dependence of the magnetization. (Symbols)
  experimental data for the nominally pure sample and the samples with
  $x=5\%$ and $x=10\%$, $\vect{H}||c$, $H=1$kOe. (Dashed lines)
  Curie laws for the $S=1/2$ $g=2.0$ paramagnets with the spin
  concentrations of 1\% and 2\% per copper ion. (Solid line)
  Reference curve illustrating expected exponential decrease of
  magnetization
  $\propto e^{-\Delta/T}$ with $\Delta=13.5$K.}\label{fig:magn}
\end{figure}

\subsection{Samples preparation and characterisation.}

Samples of pure and bromine substituted \phxc{}(PHXC for short for
the $y\neq 0$ samples) were grown from solution as described in
\cite{tatiana-growth}. The saturated solution for the growth of
pure PHCC was prepared by adding of piperazine C$_4$H$_{10}$N$_2$
(from Sigma Aldrich), dissolved in a minimal amount of
concentrated HCl, to CuCl$_2\cdot 2$H$_2$O (99.99\% purity from
Sigma Aldrich), dissolved in a minimal amount of concentrated HCl,
at a 4:1 molar ratio of CuCl$_2$ to C$_4$H$_{10}$N$_2$. Saturated
solution for the diluted crystals was prepared by introducing a
proportional amount of CuBr$_2\cdot 2$H$_2$O and HBr. We will use
nominal bromine concentration (determined by the Br to Cl ratio in
the initial solution components) for the sample identification. As
grown samples have a typical size of $3\times 3\times 5$ mm$^3$
with a well developed plane normal to the $a^*$-axis and the
longest edge of the sample parallel to the $c$-axis. Details of
the lattice structure (which is triclinic) can be found in
\cite{stone}.

The quality of the samples was checked by X-ray diffraction on a
Bruker Apex II diffractometer. It was found that up to the nominal
bromine content of 12.5\%  lattice symmetry remains the same
(triclinic) and  lattice constants slowly increase
\cite{dan-phcc-doped2}. Above 12.5\% a phase separation was found
to occur. Detailed structural analysis \cite{dan-phcc-doped2} has
demonstrated that the average bromine concentration $y$ in \phxc{}
is slightly below the nominal value $x$, $y=0.63 x$, and the
occupation of different halogen positions varies: bromine ions
enter at almost nominal concentration on the positions responsible
for the strongest couplings (rung and leg of the ladder), while
bromine occupation of the positions responsible for the
interladder couplings is by a factor of three smaller.

The static magnetization of all samples was measured with a
Quantum Design MPMS-XL system. The M(T) curves are shown on
Fig.\ref{fig:magn}. All curves demonstrate a broad maximum around
15K. Below this temperature magnetization decreases due to the
diminishing population of the gapped magnetic excitations. Since
the exchange bonds structure of PHCC is complicated (at least six
relevant couplings are suggested in Ref.\cite{stone}) we do not
attempt to fit these magnetization curves with any compact simple
model. Inelastic neutron scattering study of
Ref.\cite{prb-dan-doped} demonstrated that the gap in the spectrum
increases with doping and is above 10K for both pure and doped
samples. Thus, contribution of the thermally excited triplets to
the magnetic susceptibility at 2K is negligible (see reference
line in Figure \ref{fig:magn}). The increase of magnetization at
lowest temperatures is due to the presence of some paramagnetic
centers. A crude estimation of the amount of paramagnetic centers
can be performed by a direct comparison of the 2K magnetization
data with the Curie law. The amplitude of the magnetization at low
temperatures corresponds to the concentration of paramagnetic
centers (assuming $g=2.0$ $S=1/2$) of about 1\% per copper ion for
the nominally 5\% Br-doped sample and about 1.5\% per copper ion
for the nominally 10\% Br-doped sample (see dashed lines in Figure
\ref{fig:magn}).

\subsection{ESR experiment details}

Electron spin resonance experiments were carried out with the help
of home-made transmission type ESR spectrometers at the
frequencies 10-70 GHz and at temperatures down to 450 mK.

Limited sensitivity of our spectrometer required usage of the
quite big samples (about $3\times 3\times 5$ mm$^3$) which in turn
limited possibilities for the sample mounting. For this reason ESR
experiments for all samples were carried out for the magnetic
field $\vect{H}||a^*$ orientation. For the same reason, as the
sample size is comparable to the microwave wavelength, the field
distribution inside the sample is unknown a priori (in particular,
different polarizations of microwave magnetic field with respect
to the static field are present), but this distribution remains
the same for each series of experiments (with the same microwave
frequency and the same sample mounting).

Here we briefly recall some basics of ESR technique that will be
necessary to interpret experimental data \cite{altkoz,abblin}.

First, ESR technique allows to discern different paramagnetic
centers by their resonance fields. This allows to determine both
$g$-factor and spin of the paramagnetic center. Characteristic
fingerprint of a $S>1/2$ center is splitting of the absorption
line due to  zero-field splitting of its spin sublevels (so called
crystal field splitting which  is described in the simplest case
by axial term in the spin Hamiltonian $\hat{\cal
H}_{CF}=D\hat{S}_z^2$). Magnitude and sign of the crystal field
parameter $D$ can be deduced from the positions and relative
intensities of the split absorption components. In particular, for
a $S=1$ center a direct transition between $S_z=\pm 1$ sublevels
(so called "two-quantum" transition since it corresponds to
$\Delta S_z=\pm 2$)  results in the characteristic absorption
signal at one half of the usual paramagnetic resonance field.
Zero-field splitting of the collective triplet excitations of a
spin-gap magnet can be interpreted as  effective crystal field
microscopically originating from different anisotropic spin-spin
interactions. Unravelling the exact microscopic origin of this
splitting is a separate task, this problem has been solved by
different approaches for various spin-gap magnets in literature
\cite{zaliznyak,manaka-prb62,kolezhuk}.

Second, ESR absorption can be interpreted in absolute units. In
our experiment the microwave power transmitted through the cavity
with the sample is  recorded as a function of slowly swept
magnetic field at a fixed microwave frequency. For a weakly
absorbing point sample the decrease of the transmitted microwave
power is proportional to the imaginary part of the high-frequency
susceptibility $\chi''(\omega,H)$ for the appropriate polarization
of the microwave magnetic field. For a finite size sample the
absorption has to be integrated over the sample volume taking into
account the variation of the microwave field inside the sample.
The proportionality coefficient between $\chi''(\omega,H)$ and the
decrease of  detector output $\Delta U/U_0$ is a priory unknown,
it depends on the microwave fields distribution, cavity Q-factor
and other technical details. However, it remains the same for each
series of experiments. Thus, we have to scale our absorption with
a reference data provided by static susceptibility measurements.
If the resonance absorption spectrum consists of a single narrow
paramagnetic resonance line Kramers-Kroenig relations can be used
to scale the integrated intensity of the paramagnetic absorption
due to the microwave field perpendicular to the static field with
the static susceptibility $\chi(0)$:

\begin{eqnarray}
  \chi(0)&=&\frac{2}{\pi}\int_0^\infty \frac{\chi''(\omega)}{\omega}d\omega\approx
  \frac{2}{\pi\omega_0}\int_0^\infty\chi''(\omega)d\omega=\nonumber\\
  &=&\frac{2}{\pi H_{res}}\int_0^{\infty}\chi''(H)dH\propto\int_0^\infty\frac{\Delta
  U}{U_0}dH \label{eqn:chi-scaling}
\end{eqnarray}

\noindent This allows to establish the scaling factor for the
given experiment and to calculate the imaginary part of the
susceptibility in absolute units. Which, in turn, allows to
estimate the amount of paramagnetic centers. Note, that for the
case of $S=1$ such a scaling is applicable only to the absorption
components excited by the transversely polarised microwaves and
not to the "two-quantum" absorption.

Finally, the intensity of the ESR absorption due to  transitions
between spin sublevels $|n\rangle$ and $|m\rangle$ ($E_n<E_m$)
depends on the sublevels populations and matrix element of the
transition.

\begin{equation}\label{eqn:esr-absorption}
I_{nm} \propto
  \frac{(e^{-\frac{E_n}{T}}-e^{-\frac{E_m}{T}}){\int_s|\langle n
  |\hat{\vect{S}}\cdot \vect{h}_{mw}(\vect{r})|m\rangle|^2 dV}}{\sum_{k}e^{-E_k/T}}
\end{equation}

\noindent here the integration is performed over the sample volume
to take into account the distribution of microwave field
polarizations $\vect{h}_{mw}$ through the sample. Temperature
dependence of a selected component intensity is determined by the
combination of the Boltzmann exponents.

\section{Experimental results.}

\subsection{ESR on the reference sample $x=0$\%.}
\begin{figure}
  \centering
  \epsfig{file=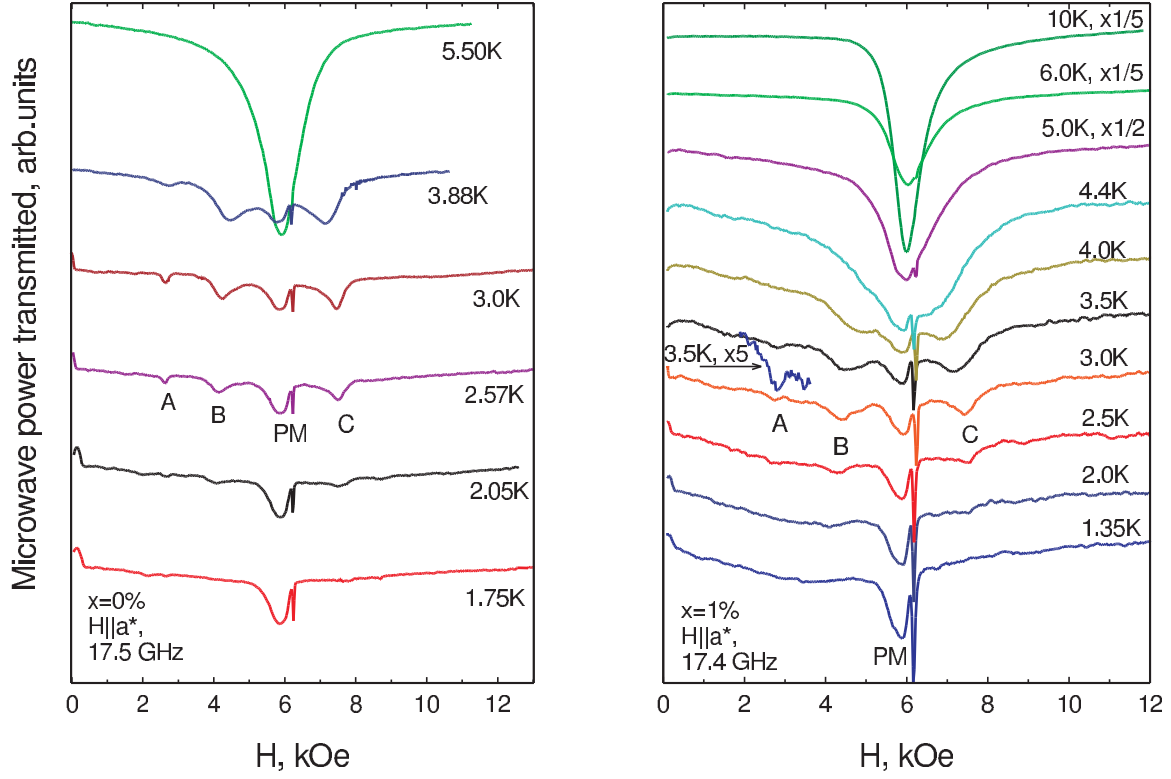, width=\figwidth, clip=}
  \caption{(color online) Temperature dependences of the ESR
  absorption in pure ($x=0$\%, left panel) and low-doped ($x=1$\%,
  right panel) samples. Spectra are offset for better
  presentation. Spectral components are marked by letters:
  (A) "two-quantum" transition, (B) and (C) main intra-triplet
  transitions, (PM) paramagnetic defects and impurities. High temperature
  spectra for the doped sample are scaled for better presentation by the factors of 1/2
  (5.0K) and 1/5 (6.0K and 10K), "A" component of the 3.5K spectra
  for the doped sample is magnified by the factor of 5 for better presentation.
  Narrow absorption line at 6.2kOe is a DPPH marker
  ($g=2.0$).}\label{fig:low-17ghz-scans}
\end{figure}

The results of  ESR investigation of the nominally pure ($x=0$\%)
PHCC are reported in detail in \cite{glazkov-phcc}: At high
temperatures (above 20K) a single resonance absorption component
is observed with an anisotropic $g$-factor slightly above 2.0, as
it is typical for Cu$^{2+}$ ions. Below 20K the ESR absorption
loses intensity as the magnetic excitations freeze out. As the
population of magnetic excitations drops down, their interaction
vanishes and the zero-field splitting of the triplet $S=1$
excitations appears.

In Figure  \ref{fig:low-17ghz-scans} we present reference data
 for the $x=0$\% sample taken at the same frequency and
orientation as the data for the bromine-diluted samples. As was
described above, there is a single component ESR absorption line
above 5K. The ESR absorption line splits around 4K into three
components: two main components (Fig.\ref{fig:low-17ghz-scans})
corresponding to $|\pm 1\rangle\leftrightarrow|0\rangle$
transitions and the weaker component   at approximately half of
the paramagnetic resonance field corresponding to the
$|+1\rangle\leftrightarrow|-1\rangle$ transition. We mark these
components in the order of increasing resonance field as "A", "B"
and "C" corresponding to the "two-quantum" transition and the two
main components. Splitting of the main components and their
relative intensities are related to the parameters of the
effective crystal field and to the orientation of the anisotropy
axes with respect to the field. Intensity of the "C" component
slightly exceeds the intensity of the "B" component. Split
components continue to lose intensity on cooling and below 2.0K
the ESR absorption spectrum mainly consists of the paramagnetic
absorption due to the defects and impurities (marked as "PM").

\subsection{ESR on the sample with low nominal bromine content $x=1$\%.}

Temperature evolution of ESR absorption for the sample with low
nominal bromine content ($x=1$\%) is qualitatively the same as for
the pure sample (Fig.\ref{fig:low-17ghz-scans}). On cooling below
10K the ESR absorption loses intensity, the ESR line broadens and
splits at approximately 4K into three components (again labelled
as "A", "B" and "C" on Fig.\ref{fig:low-17ghz-scans}). These
components freeze out on further cooling and at the lowest
temperature the ESR absorption spectrum consists of  single line
with  $g$-factor close to 2.0, which is related to defects and
impurities. Thus, the observed ESR absorption in the low-doped
sample is mostly due to the triplet excitations, as in the pure
sample. As in the case of  pure compound, intensity of the "C"
component for the 1\% doped sample slightly exceeds that of the
"B" component, which indicates that the effective anisotropy
constant is also with the same sign.

\subsection{ESR on the   samples with high  nominal bromine content $x=$5 and 10 \%.}
\begin{figure}
  \centering
  \epsfig{file=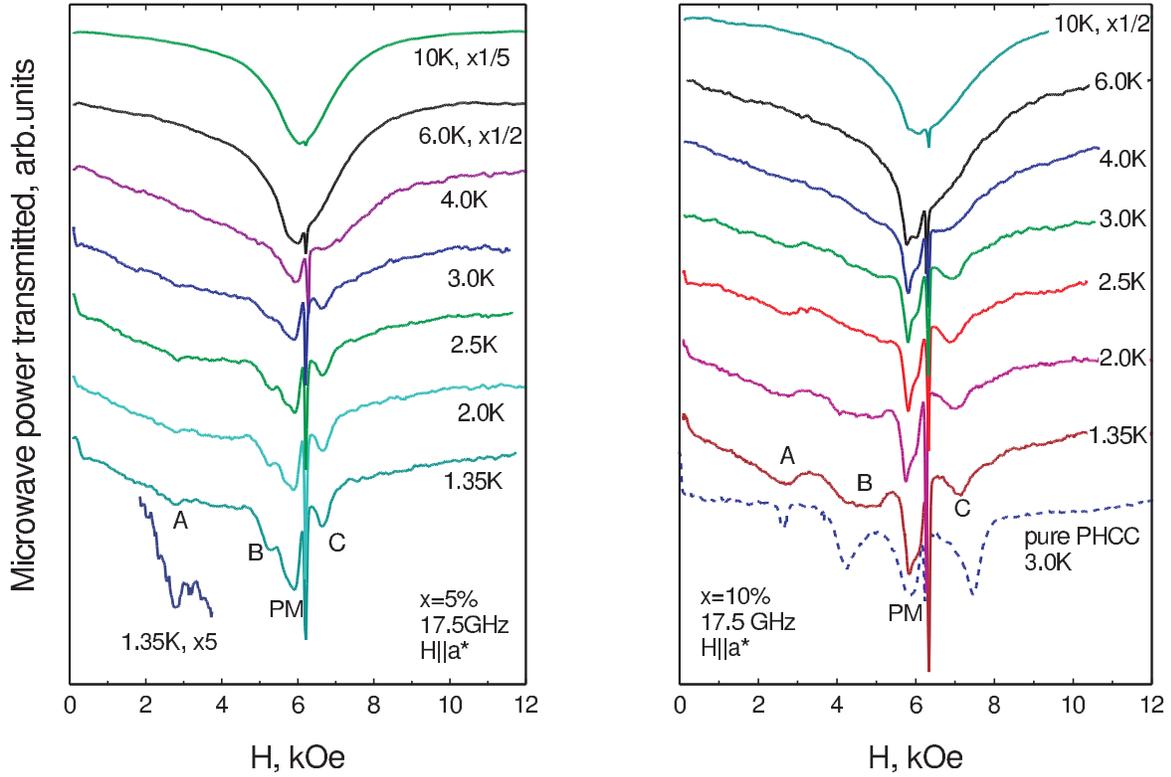, width=\figwidth, clip=}
  \caption{(color online) ESR absorption spectra for the PHXC with high nominal bromine content
 $x=$5\% (left panel) and $x=10$\% (right panel).
 Spectra are offset for better presentation. Spectral components are marked by letters
  in the same way as for the pure compound. High temperature
  spectra are scaled for better presentation: for the $x=5$\% sample by the factors of 1/2
  (6.0K) and 1/5 (10K) and for the $x=10$\% sample by the factor
  of 1/2 (10K), "A" component of the 1.35K spectra for $x=5$\% sample is amplified by
  the factor of 5 for better presentation. Narrow
  line at 6.2kOe is a DPPH marker ($g=2.00$). Dashed line at the
  right panel shows unscaled 3.0K ESR absorption in pure
  PHCC.}\label{fig:high-17ghz-scans}
\end{figure}

\begin{figure}
  \centering
  \epsfig{file=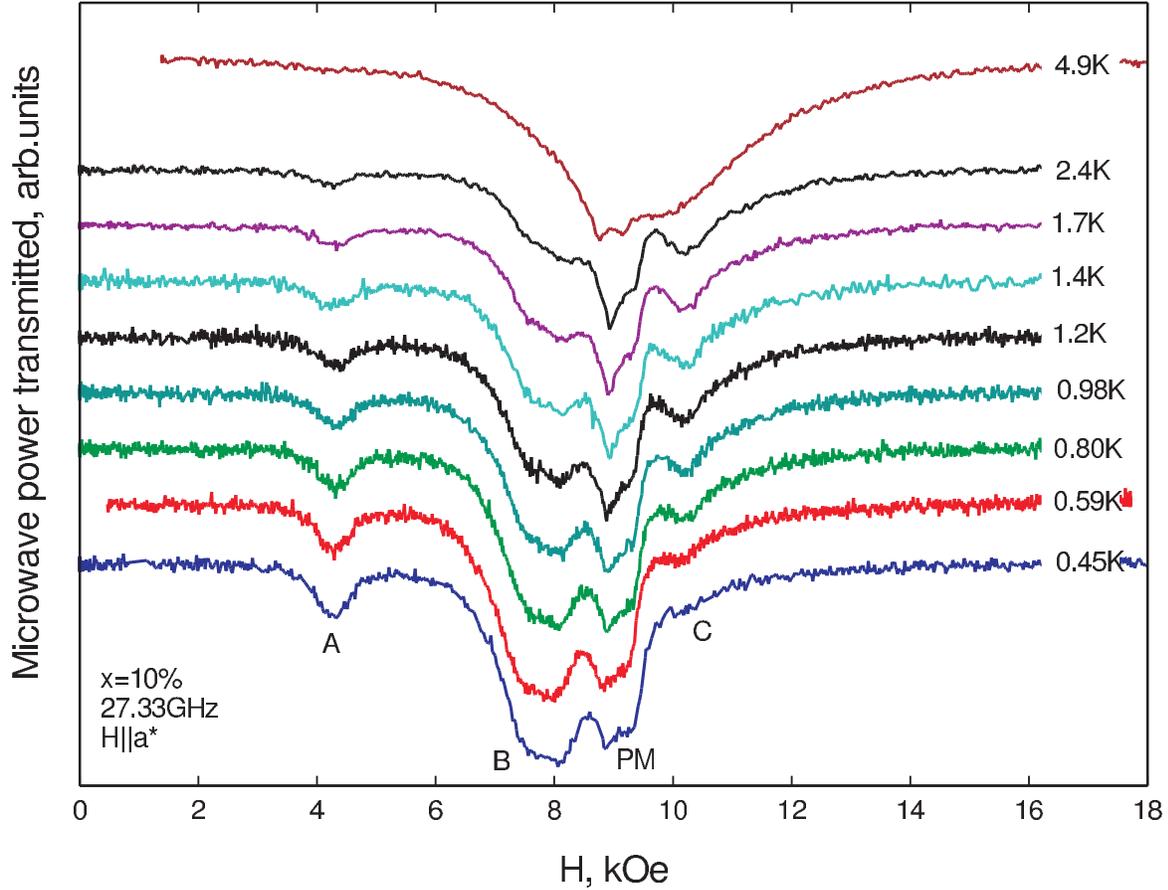, width=\figwidth, clip=}
  \caption{(color online) ESR absorption spectra for the PHXC with nominal Br
  concentration $x=$10\% below 1K. The spectra are offset for better presentation. Spectral components are marked by letters
  in the same way as for the pure compound.
  $\vect{H}||a^*$, $f=27.33$ GHz.}\label{fig:10-he3-scans}
\end{figure}

\begin{figure}
  \centering
  \epsfig{file=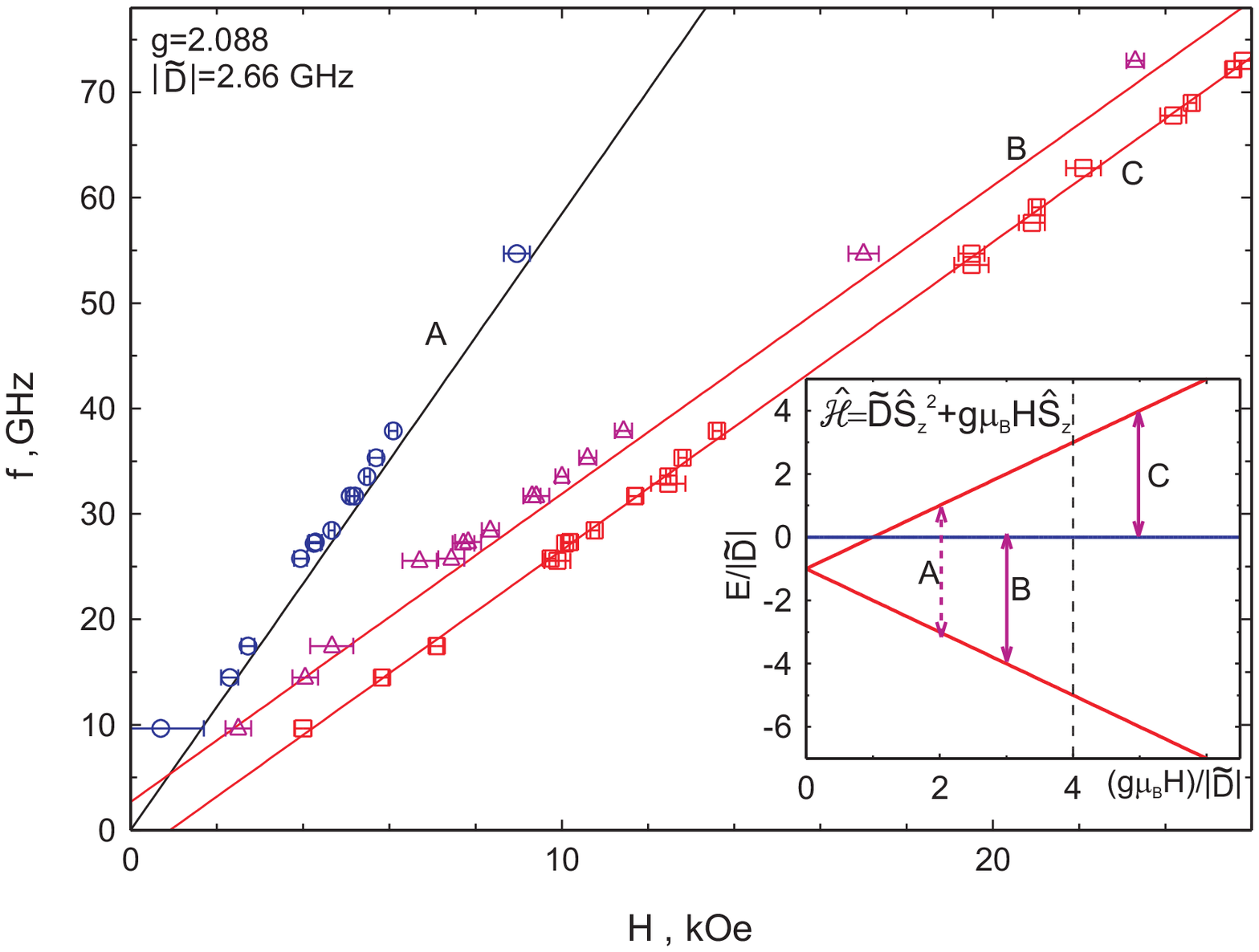, width=\figwidth, clip=}
  \caption{(color online) Frequency-field dependences for the PHXC with nominal Br
  concentration $x=$10\%. Spectral components are marked by letters
  in the same way as for the pure compound. Solid lines are model
  curves for the $S=1$ in the axial crystal field.
  $\vect{H}||a^*$, $T=1.4-3.0$ K. (Inset) Scheme of the field
  dependence of the energy levels for the $S=1$ center in the axial
  crystal field for the case of negative anisotropy constant
  $\widetilde{D}$. Vertical arrows of the same length mark positions of
  resonance fields in the experiment with fixed microwave
  frequency. Vertical dashed line marks position of the resonance field in the case of $\widetilde{D}=0$.
  Different resonance transitions are labelled "A", "B'
  and "C" in the same way as the absorption  components}\label{fig:10-f(h)}
\end{figure}

Temperature evolution of ESR absorption in the samples of PHXC
with higher nominal bromine concentration $x=$5\% and 10\%
(Fig.\ref{fig:high-17ghz-scans}) is qualitatively different from
that of the lightly doped sample. At  temperatures above 4 K the
intensity of absorption decreases with cooling and the splitting
of the ESR line into similar components occurs. But the split
components remain clearly visible even at the lowest temperatures.
Below we will focus our consideration on the sample with $x=10$\%,
absorption spectra for the sample with $x=5$\% is qualitatively
similar to the $x=10$\% case.

To check whether this absorption signal corresponds to the $S=1$
objects or not and to follow its temperature evolution, we
performed measurements at temperatures down to 450mK on the
$x=10$\% sample and measured the frequency-field dependence of the
observed ESR absorption signal. These data demonstrate (Fig.
\ref{fig:10-he3-scans}) that  "A" and "B" components of the split
absorption spectrum survive down to the lowest temperatures and
gain intensity on cooling, while "C" component loses intensity
below 1K and almost disappears at 450 mK. The frequency-field
dependences (Fig. \ref{fig:10-f(h)}) for all three modes are in
quantitative agreement with  known \cite{altkoz,abblin}
frequency-field dependences for a $S=1$ object  in an axial
crystal field with  $g$-factor $g=2.088$ and effective anisotropy
constant $|\widetilde{D}|=2.7\pm0.5$ GHz.

Contrary to the case of pure and low-doped samples, the integrated
intensity of  "B" component at $T>1.35$K (Fig.
\ref{fig:high-17ghz-scans}) slightly exceeds the intensity of the
"C" component, this difference becomes obvious on cooling below 1K
(Fig. \ref{fig:10-he3-scans}). Thus, the effective anisotropy
constant in $\vect{H}||a^*$ orientation for the high-doped samples
has a different sign from that for the pure or low-doped samples.

\section{Discussion.}
\subsection{Effect of doping on the paramagnetic resonance at $T=10$K.}
\begin{figure}
  \centering
  \epsfig{file=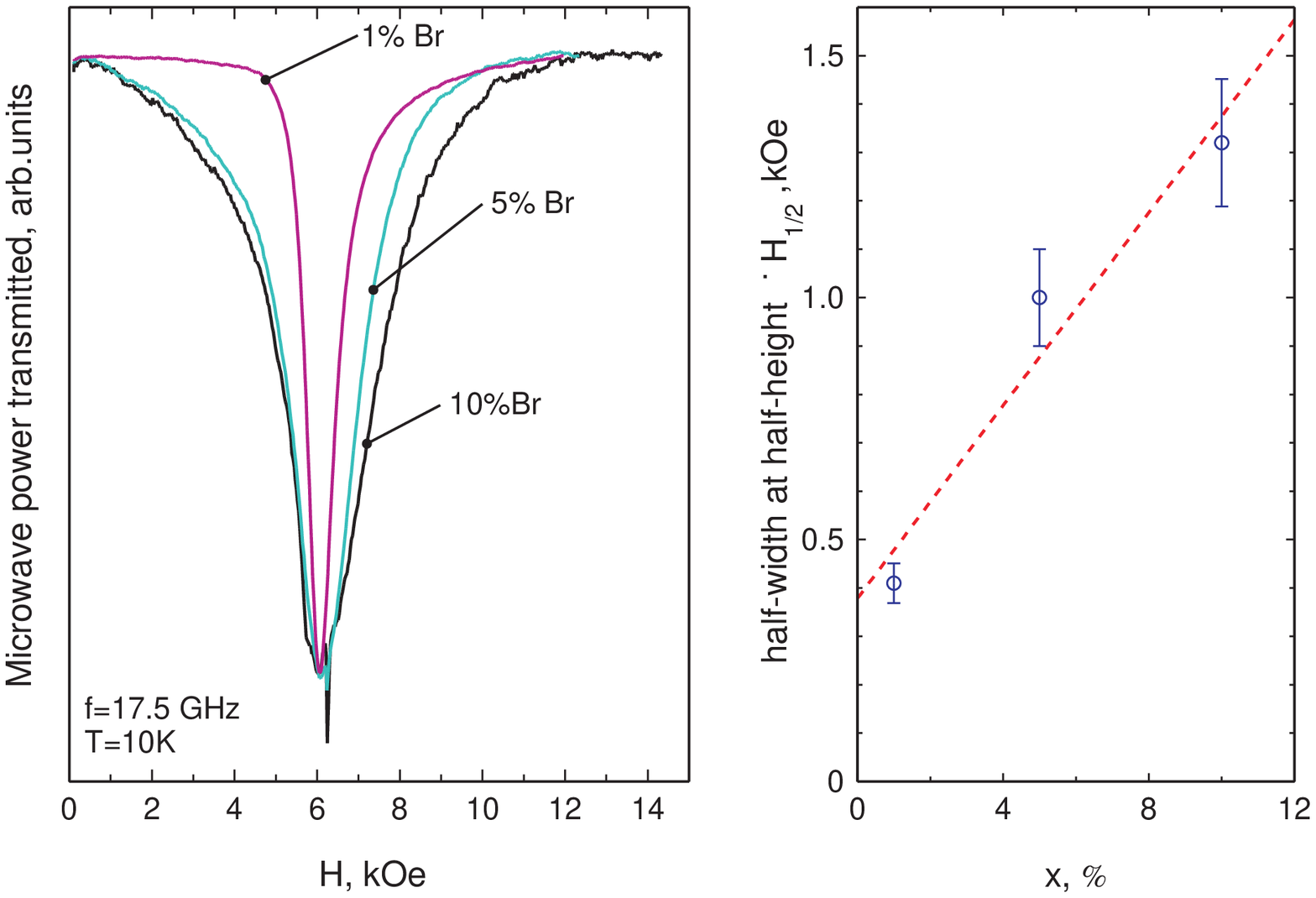, width=\figwidth, clip=}
  \caption{(color online) ESR  absorption spectra at different doping levels (left) and
  concentration dependence of the half-width at half-height
  (right). ESR absorption spectra are scaled to the same amplitude
  for better presentation, narrow absorption line at 6.2kOe is a
  DPPH marker ($g=2.00$), dashed line on the right panel is a
  guide to the eye. $f=17.5$GHz, $T=10$K. }\label{fig:10k-width}
\end{figure}

At temperatures above 6K ESR absorption line for all the samples
consists of a single component and demonstrates an increase of the
intensity on heating. This increase of intensity is in an
agreement with the increase of static susceptibility on heating in
the same temperature range. Thus, ESR absorption at these
temperatures is dominated by the triplet excitations of the
spin-gap magnet. Comparison of  ESR absorption lines measured at
the same temperature $T=10$K (Figure \ref{fig:10k-width})
demonstrates a monotonous increase of  linewidth with doping. As
ESR linewidth is directly related to the lifetime of the spin
precession, i.e. to the lifetime of the triplet excitations, this
observation is in qualitative agreement with  increase of magnon
damping rate observed in neutron scattering experiments.
\cite{prb-dan-doped,prb-dan-doped-magnons}

\subsection{Estimation of the amount of paramagnetic defects in doped samples.}

The "PM" mode observed in all samples with $g$-factor close to 2
increases in intensity on cooling and has an irregular lineshape.
This allows to interpret it as being related to some uncontrolled
paramagnetic defects (e.g., off-site copper ions, paramagnetic
impurities etc.). To  estimate the concentration of the
paramagnetic centers we scale the integral intensity of the "PM"
mode measured at 17.5GHz (Figures \ref{fig:low-17ghz-scans},
\ref{fig:high-17ghz-scans}) at low temperatures (between 1.5K and
approximately 3K) with the total intensity of the ESR signal at
high temperature (10K). The high temperature total intensity can
then be scaled to the measured static magnetization (see
Eqn.(\ref{eqn:chi-scaling})). Then the concentration of the
paramagnetic centers can be estimated from the Curie law. We
neglect here anisotropy of the PHCC $g$-factor which would yield
insignificant corrections \cite{g-comment}.

We have found that the intensity of the "PM" mode in all the
samples corresponds to approximately 0.1\% $S=1/2$ $g=2.1$
(measured $g$-factor) paramagnetic centers per copper ion (Table
\ref{tab:pm}).

\begin{table}
\caption{Estimated amount of paramagnetic defects (assuming
$S=1/2$ $g=2.1$) in different samples.}\label{tab:pm}
\centering
\begin{tabular}{|c|c|}
\hline
 Nominal Br & Amount of paramagnetic\\
 concentration, \% & centers contributing to "PM" \\
 &component, \% per copper\\
\hline
     0&$0.09\pm0.01$\\
     1&$0.09\pm0.01$\\
     5&$0.06\pm0.01$\\
     10&$0.14\pm0.03$\\
\hline
\end{tabular}
\end{table}
Apparent  increase of  amplitude of the "PM" mode with increasing
bromine concentration in Figures
\ref{fig:low-17ghz-scans},\ref{fig:high-17ghz-scans} is  purely
visual effect due to broadening of the main absorption line.

Note that the low-temperature magnetization  of the doped samples
(Figure \ref{fig:magn}) corresponds to  ten-fold higher amount of
paramagnetic centers. This discrepancy with ESR observations rules
out trivial creation of paramagnetic defects on doping.
Additionally, the estimated amount of the paramagnetic defects is
much smaller than the nominal Br concentration. This implies that
Br substitution indeed affects the magnetic bonds instead of
creating additional structural or magnetic defects in agreement
with X-ray analysis of Ref.\cite{dan-phcc-doped2}.

\subsection{Nature of the $S=1$ resonance centers in the
samples with high bromine contents.}

\begin{figure}
  \centering
  \epsfig{file=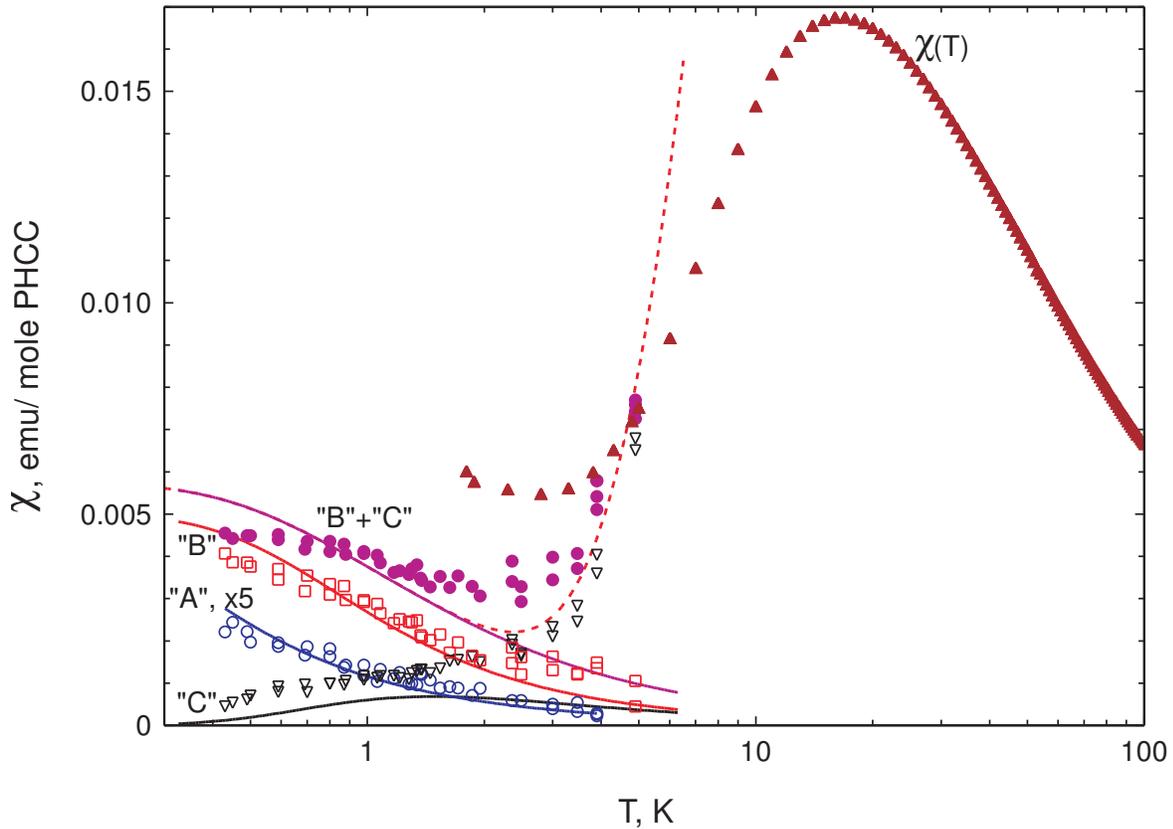, width=\figwidth, clip=}
  \caption{(color online)Temperature dependences of the ESR
  absorption intensities at low temperatures scaled with the
  high-temperature magnetization for the PHXC sample with nominal Br
  concentration $x=10$\%. (open circles) Intensity of the "A"
  component amplified by the factor of 5.0 for better
  presentation; (open squares) intensity of the "B" component;
  (open triangles) intensity of the "C" component; (closed
  circles) total intensity of the "B" and "C" components; (closed
  triangles) static susceptibility. Curves are fits as described
  in the text.}\label{fig:he3-i(t)}
\end{figure}

The ESR absorption spectra measured at the  lowest temperatures
(Fig. \ref{fig:10-he3-scans}) prove that the low-temperature
absorption in the high-doped samples is dominated by some other
$S=1$ objects affected by the effective crystal field $\hat{\cal
H}_{CF}=\widetilde{D} \hat{S}_z^2$ (here we assume $z||a^*$).The
magnitude of the constant $\widetilde{D}=(-2.7\pm0.5)$~GHz and the
$g$-factor value $g=2.088$ are determined from the frequency-field
dependence $f(H)$, see Figure \ref{fig:10-f(h)}. The negative sign
of the $\widetilde{D}$  is evidenced by the higher relative
intensity of the "B" component as this transition corresponds to
the excitation from the lowest sublevel (see insert of Figure
\ref{fig:10-f(h)}).

Temperature dependence of the ESR absorption intensity (Figure
\ref{fig:he3-i(t)}) was determined for components "A", "B" and "C"
by fitting them with a Lorentzian lineshape. Determination of
intensity of mode "A" is the most reliable since this mode is well
separated from other absorption components. But mode "A" can not
be used for the determination of the $S=1$ objects concentration
because of the unknown excitation conditions for this
"two-quantum" transition.  The modes "B" and "C" are excited by
the conventional transverse polarization and their intensity can
be scaled with the static susceptibility data. Closeness of the
modes "B" and "C" with the defects contributed mode "PM"
complicates fit procedure, leading in particular to the systematic
overestimation of the intensity of the weaker mode "C" located on
the shoulder of the "PM" mode. However, as the scaling with the
static magnetization is done at 5K where the "PM" mode intensity
is relatively weak and the low-temperature (below 0.6K) absorption
is dominated by the mode "B", the absolute value of ESR absorption
at low temperatures (which is determined by the concentration of
the $S=1$ objects) can be determined reliably. When scaling the
ESR absorption with the static susceptibility we neglect the
$g$-factor anisotropy which would yield only small (about 5\%)
correction in our case \cite{g-comment}.

To model temperature dependence of the ESR intensity  we assumed
that these $S=1$ objects are gapless with energy of the
field-independent $S_z=0$ substate being zero. Temperature
dependences of ESR intensities can be then deduced from
Eqn.\ref{eqn:esr-absorption} (see Appendix for details) with known
microwave frequency, $g$-factor and anisotropy constant, leaving
the scaling factors as the only fitting parameters. Experimentally
measured intensities of the ESR modes follow modelled curves well.
Best fit is obtained with the concentration of the $S=1$ objects
$x_{S=1}=0.004\pm 0.001$ per molecule of PHXC . The results of
this fit are shown on Figure \ref{fig:he3-i(t)}.

At higher temperatures (above 4K) the observed ESR absorption is
dominated by  thermally excited triplets. Temperature dependence
of the total ESR absorption intensity in the temperature range
from 0.45K to 6K can be fitted by a sum of  contributions from the
$S=1$ objects and a gapped contribution $b\times e^{-\Delta/T}$
with the gap value fixed to $\Delta=15$K. The gap value was
obtained by  extrapolation of the neutronographic data
\cite{prb-dan-doped,prb-dan-doped-magnons} (see dashed line at
Figure \ref{fig:he3-i(t)}). Also, note that as temperature
increases, the dominating ESR component slowly switches from the
low-field "B" component to the higher-field "C" component. This
redistribution of absorption intensities follows crossover from
the  absorption dominated by gapless triplets at low temperatures
to the  absorption dominated by triplet excitations  at higher
temperatures.

The origin of these $S=1$ gapless triplets can not be
unambiguously recovered from our data. However, the finding that
concentration of these objects is close to 1\% per PHXC molecule
for the 10\% nominal doping suggests the following explanation.
The building block of PHCC magnetic subsystem is a Cu$_2$Cl$_6$
dimer. Two of the chlorine ions are responsible for the intradimer
coupling and the four remaining chlorines participate in the
formation of the inter-dimer couplings. At the bromine doping
level of 10\% (assuming uniform doping) the probability to find a
dimer with both intra-dimer bond-forming halogens being bromine
ions is exactly 1\%. A detailed structural analysis of
\cite{dan-phcc-doped2} suggested that the occupation of the
inter-dimer halogen positions (position Cl$_1$ in terms of
\cite{dan-phcc-doped2}) is slightly less then the nominal value
and is about 7.5\%, which makes the probability of the double
substitution 0.56\% even closer to our estimation of the $S=1$
objects concentration. Thus, dimers with two of the intra-dimer
exchange bonds modified by bromine substitution are the possible
source of the gapless $S=1$ triplets. These triplets could be
interpreted as a localized triplet mode, created by a potential
well centered on such a doubly substituted dimer. In the limit of
strong localization this localized triplet mode transforms into a
ferromagnetically coupled pair of copper spins.

Finally, we would like to note that the contribution of the $S=1$
spin to the susceptibility exceeds that of the $S=1/2$ spin  by
the factor of 8/3 (spin enters Curie law as $S(S+1)$). Thus, the
determined concentration of 0.4\% of $S=1$ objects per PHCC
molecule (or 0.2\% per copper ion) corresponds to the
concentration of about 0.5\% of $S=1/2$ centers per copper ion
which is reasonably close to the amount estimated from the
magnetization data (Figure \ref{fig:magn}). Additional
contribution to the static magnetization of the doped PHXC should
also arise from a Van-Vleck-like mechanism suggested in
\cite{dan-phcc-doped2}.

\section{Conclusions.}

We have studied the effect of bond-doping on the spin-gap magnet
\phcc. Our experiments show that substitution of  chlorine ions by
bromine  up to 10\% nominal doping does not lead to the
destruction of the spin gap state. We have shown, that
introduction of bromine ions does not lead to  formation of
structural defects
--- the concentration of  paramagnetic $S=1/2$ centers remains
practically the same (about or below 0.1\%) for the full set of
samples ranging from  pure compound to 10\% nominally doped
sample.

However, bond-doping does affect thermally excited triplet
excitations by shortening their lifetime, as evidenced by increase
of ESR linewidth.

Finally, we have established that the low-temperature ESR
absorption is dominated by gapless $S=1$ triplets formed with
doping.

\section{Acknowledgements.}

The work was supported by the Russian Foundation for Basic
Research (RFBR) projects No.12-02-31220, No.12-02-00557 and
No.15-02-05918, Russian Presidential Grant for the Support of the
Leading Scientific Schools No.4889.2012.2 and No.5517.2014.2. This
work was partially supported by the Swiss National Fund (division
2). This project was partially supported by the Estonian Ministry
of Education and Research under grant IUT23-03 and Estonian
Research Council grant  PUT451.

\appendix
\section{Equations for the temperature dependences of ESR modes
intensities}

If we assume that the $S=1$ objects responsible for the
low-temperature ESR absorption have no energy gap, then, taking
energy of the field independent $S_z=0$ sublevel for zero,
energies of the $S_z=\pm 1$ sublevels are $E_{\pm
1}=\widetilde{D}\pm g\mu_B H$. The temperature dependence of the
intensity for each of the resonance absorption components is
described by the combination of Boltzmann exponents
(Eqn.(\ref{eqn:esr-absorption})).

For the "two-quantum" transition "A" this yields

\begin{equation}
  I_A \propto \frac{e^{-\frac{\widetilde{D}-g\mu_B H}{T}}-e^{-\frac{\widetilde{D}+g\mu_B H}{T}}}{1+e^{-\frac{\widetilde{D}-g\mu_B H}{T}}+e^{-\frac{\widetilde{D}+g\mu_B
  H}{T}}}=e^{-\frac{\widetilde{D}}{T}} \frac{sinh\frac{h\nu}{2T}}{2+e^{-\frac{\widetilde{D}}{T}}cosh\frac{h\nu}{2T}}
\label{eqn:ESRintensity-A}
\end{equation}

The main absorption components "B" and "C" correspond to the
ordinary $\Delta S_z=\pm 1$ transitions and are excited by the
microwave magnetic field perpendicular to the static field. This
allows to analyse the intensities of these modes quantitatively
and to estimate the concentration of these gapless $S=1$ objects.
The scaling factor between the observed absorption intensity and
the static susceptibility was calculated at  temperatures from 5
to 6 K, which are the highest temperatures available in our He-3
experimental setup. At these temperatures the intensity of the
absorption signal related to the paramagnetic defects is
negligible compared to the total intensity. This allows to express
the intensities of  "B" and "C" components in terms of
concentration of the $S=1$ objects:

\begin{eqnarray}
  I_B&=&x W \frac{e^{-\frac{\widetilde{D}-g\mu_B H}{T}}-1}{1+e^{-\frac{\widetilde{D}-g\mu_B H}{T}}+e^{-\frac{\widetilde{D}+g\mu_B
  H}{T}}}=x W \frac{e^{\frac{h \nu}{T}}-1}{1+e^{\frac{h\nu}{T}}+e^{-\frac{2\widetilde{D}}{T}}e^{-\frac{h\nu}{T}}}
\label{eqn:ESRintensities-B}\\
  I_C&=&x W \frac{1-e^{-\frac{\widetilde{D}+g\mu_B H}{T}}}{1+e^{-\frac{\widetilde{D}-g\mu_B H}{T}}+e^{-\frac{\widetilde{D}+g\mu_B
  H}{T}}}=x W \frac{1-e^{-\frac{h\nu}{T}}}{1+e^{-\frac{h\nu}{T}}+e^{-\frac{2\widetilde{D}}{T}}e^{\frac{h\nu}{T}}}
\label{eqn:ESRintensities-C}
\end{eqnarray}

The scaling factor $W$ ensures the correct Curie dependence of the
total intensity at high temperatures ($T\gg \widetilde{D},g\mu_B
H, h\nu$):
\begin{equation}\label{eqn:ht-limit}
  I_{tot}=I_B+I_C\approx \frac{2}{3} xW
  \frac{h\nu}{T}=x \frac{g^2\mu_B^2N_A S(S+1)}{3T}
\end{equation}

\noindent for our data $W=g^2\mu_B^2N_A/(h\nu)\approx
1.25$emu/(mole PHXC).


\begin{thebibliography}{10}
\bibitem{g-comment} Static susceptibility was measured for
$\vect{H}||c$, while the ESR absorption is recorded for
$\vect{H}||a^{*}$. The absorption by thermally activated triplets,
"B", "C" and "PM" modes are excited by the transversely polarized
microwaves with the microwave magnetic fields distributed in the
$(bc)$-plane. Due to the big sample size, various polarizations
are present. Value of the $g$-factor varies in the $(bc)$-plane
from 2.05 to 2.14 \cite{glazkov-phcc}. Accounting for this
$g$-factor variation would give a correction of about 5\% for the
scaling between the static susceptibility and the integrated ESR
absorption , which is much less then the accuracy of the
concentration determination.
\end{thebibliography}
\end{document}